\relax

\documentclass[runningheads]{llncs}

\usepackage{amsmath}
\DeclareMathOperator*{\argmax}{{\rm argmax}}
\usepackage{graphicx}
\usepackage{algorithm}
\usepackage[noend]{algpseudocode}

\newcommand{\citep}{\cite}

\begin{document}
 
\title{Stackelberg Punishment and \\ Bully-Proofing Autonomous Vehicles}

\author{Matt Cooper \and
Jun Ki Lee \and
Jacob Beck \and
Joshua D. Fishman \and
Michael Gillett \and
Zo\"{e} Papakipos \and
Aaron Zhang \and
Jerome Ramos \and
Aansh Shah \and
Michael L. Littman}
\authorrunning{M.\ Cooper et al.}
%
\institute{Brown University, Providence, RI 02912, USA\\
\email{matthew\_cooper@alumni.brown.edu}\\
\url{https://cs.brown.edu}}


\maketitle
\begin{abstract}
Mutually beneficial behavior in repeated games can be enforced via the threat of punishment, as enshrined in game theory's well-known ``folk theorem.'' There is a cost, however, to a player for generating these disincentives. In this work, we seek to minimize this cost by computing a ``Stackelberg punishment,'' in which the player selects a behavior that sufficiently punishes the other player while maximizing its own score under the assumption that the other player will adopt a best response. This idea generalizes the concept of a Stackelberg equilibrium.
Known efficient algorithms for computing a Stackelberg equilibrium can be adapted to efficiently produce a Stackelberg punishment. We demonstrate an application of this idea in an experiment involving a virtual autonomous vehicle and human participants. We find that a self-driving car with a Stackelberg punishment policy discourages human drivers from bullying in a driving scenario requiring social negotiation.
\end{abstract}


\section{Introduction}

Driving is an inherently social activity, and it will remain so as long as human drivers share the road with autonomous vehicles. The social nature of driving introduces several problems that are often overlooked in self-driving car (SDC) research. An effective SDC will have to account for the variance in human driving styles and preferences~\citep{basu17}, as well as varying norms~\citep{bruce05} and laws~\citep{brodsky16} in different parts of the world. Additionally, SDCs will have to navigate the many nuanced ``corner cases'' of driving that require complex social negotiation.

Current planning algorithms for self-driving cars are hard-coded and programmed to be cautious. Caution is important for safety, but single-minded caution can make it impossible to complete required driving tasks. For example, to merge onto a busy highway, a driver must stick the car's nose out and encourage other drivers to slow down~\citep{chesterman16}. Furthermore, human drivers can take advantage of overly-cautious SDCs by driving aggressively, effectively bullying SDCs by forcing them to yield when they should otherwise have the right of way. Such behavior has been noted as a possible impediment to mainstream adoption of SDCs~\citep{tennant16,brooks17}.

In this work, we explore a computational game theoretic approach that might be useful in these scenarios. In particular, we examine the idea of using a strategy that adaptively discourages anti-social behavior while remaining safe. Our proposed strategy has the overall structure of the ``folk theorem'' of repeated games---stabilize mutually beneficial behavior with the threat of punishment in later rounds of play~\citep{osborne94}. However, unrestricted punishment could include unsafe behavior like intentionally crashing into the opponent's car. We propose using a punishment strategy that only restricts the opponent's utility to some safe target level while maximizing the utility of the agent. With analogy to a Stackelberg equilibrium, which is the strategy with maximum utility when paired with the opponent's best response, we call such a strategy a \emph{Stackelberg punishment}.


A Stackelberg punishment can be computed efficiently in several classes of games for which efficient Stackelberg equilibria algorithms exist. We demonstrate the concept deployed in a simple but strategically relevant driving scenario of negotiating right of way on a one-lane bridge.


In the first part of this paper,
we discuss efficient algorithms for computing Stackelberg punishment. In the second part, we describe an application of a Stackelberg punishment to solving the SDC bullying problem and demonstrate its efficacy in an experiment with human participants.

\section{Tree-Based Games}

We define a tree-based alternating-move two-player game as a tuple $\langle \mathcal{S}, \mathcal{A}, \mathcal{B}, I, \mathcal{F}, \\s_r, T, R \rangle$. Here, $\mathcal{S}$ is a set of states with $\mathcal{A} \subseteq \mathcal{S}$ being the subset of states where the first player (the leader) has control, $\mathcal{B} \subseteq \mathcal{S}$ being the subset of states where the second player (the follower) has control, and $\mathcal{F} \subseteq \mathcal{S}$ being the subset of states that are final states (leaves), ending the decision process. The sets $\mathcal{A}$, $\mathcal{B}$, and $\mathcal{F}$ partition $\mathcal{S}$. The initial state $s_r$ is the root of the tree. The set $I$ is the set of actions available at each state with the transition function $T: \mathcal{S} \times I \rightarrow \mathcal{S}$ returning the next state reached from non-final state $s \in \mathcal{S}\setminus \mathcal{F}$ when action $i$ is selected. The state space forms a tree in that each state can be reached by only one path from the root. The pair $R(s) = (\alpha,\beta)$ is the reward values obtained by the players when state $s\in\mathcal{F}$ is reached.

A policy $\pi(s,i)$ maps each non-final state $s\in \mathcal{S}\setminus\mathcal{F}$ and action $i$ to the probability that action $i$ is taken in that state. We can define the value of a policy $\pi$ from state $s\in \mathcal{S}\setminus\mathcal{F}$ as 
$$V^\pi(s) = \sum_i \pi(s,i) V^\pi(T(s,i)),$$
where $V^\pi(s) = R(s)$ for $s\in\mathcal{F}$. That is, $V^\pi(s)$ represents the payoff pair for the two players if they adopt the joint (stochastic stationary Markov) strategy $\pi$. The leader makes the selection in the $\mathcal{A}$ states and the follower makes the selection in the $\mathcal{B}$ states. Given a policy $\pi_A$ defined on the states in $\mathcal{A}$ and a policy $\pi_B$ defined on the states in $\mathcal{B}$, we write $\pi = (\pi_A,\pi_B)$ to represent the policy $\pi(s,i) = \pi_A(s,i)$ if $s\in\mathcal{A}$ and $\pi_B(s,i)$ if $s\in\mathcal{B}$.

For policies for the two players, $\pi_A$ and $\pi_B$, we can write $V_A(\pi_A,\pi_B)$ and  $V_B(\pi_A,\pi_B)$ representing the expected payoffs to the two players when these policies are executed:
$$(V_A(\pi_A,\pi_B), V_B(\pi_A,\pi_B))
 = V^\pi(s_r).$$
Given such a tree-based game and a policy $\pi_A$, we call a policy $\pi_B$ a \emph{best response} if, for all $\pi_B'$, $V_B(\pi_A,\pi_B) \geq V_B(\pi_A,\pi_B')$. That is, the follower cannot improve its value by adopting a different policy. We write $\pi_B = M(\pi_A)$ for the best response to $\pi_A$.

In this setting, a Stackelberg equilibrium policy for the leader is
$$\argmax_{\pi_A} V_A(\pi_A,M(\pi_A)).$$
That is, assuming the follower will adopt a best response to whatever the leader elects to do, the leader behaves so as to maximize its reward.

Letchford and Conitzer~\cite{letchford10} introduced an efficient algorithm for computing Stackelberg equilibria in tree-based games. In a tree with $n$ leaves and $m$ internal nodes, their approach runs in time $O(m n^2)$. The algorithm works by determining, for each state $s$, a set of payoff pairs that can be obtained through some choice of $\pi_A$ and a best response $\pi_B$. Since the objective is to find a policy $\pi_A$ that maximizes reward for the leader, we need only maintain the points of this set that maximize reward for the leader for each possible value obtained by the follower. 

The algorithm represents the set via a finite set of payoff points $P(s)$ and a finite set of line segments $L(s)$ connecting some subset of the points in $P(s)$. It builds up the representation for a state $s$ out of the representation computed for the children of $s$. At the leaves of the tree, the representation is simply the rewards at the leaves. 

For a state $s$ where the leader selects the action, the representation is computed by noting that the leader can choose any child of $s$, therefore any of the achievable payoffs at any of the children of $s$ are also achievable. However, the leader can also probabilistically select any of its children. This translates into line segments where one endpoint comes from the representation of one child node and the other endpoint comes from the representation of a different child node. This set is sufficient for capturing the representation of the set of possible values at $s$, but it may include some unnecessary lines (or even points). These extra bits of representation can be removed or ignored, as we are ultimately only concerned with the points with maximum value for the leader.

For a state $s$ where the follower selects the action, the follower will select the action that gives it the highest value, assuming it has adopted a best response policy. For an action $i$, we compute $\sigma_i$ to be the lowest value for which one should be willing to tolerate selecting $i$ over the alternatives. To compute this value, we assume the leader will make the alternatives maximally unattractive. We then modify the points and lines representing the child's values to reflect this preference. Once that modification is completed, every value for every child can be achieved at $s$.

Once the set of points and lines needed to represent the achievable values at the root $s_r$ are computed, the point with the largest value for the leader can be returned as the value of the Stackelberg equilibrium for the tree. (Computing the policy itself involves unrolling this computation in reverse order and is detailed in the original paper.)


A single change is all that is needed to adapt the algorithm to produce a Stackelberg punishment---the strategy must also result in an expected reward for the follower that does not exceed $\theta$:
$$\argmax_{\pi_A: V_B(\pi_A,M(\pi_A)) \leq \theta} V_A(\pi_A,M(\pi_A)).$$
That is, the leader maximizes its reward against a best responding follower while holding the follower's value to a cap of $\theta$. It follows that the leader cannot improve its value without the follower's value rising above $\theta$. 

Our algorithm for computing a Stackelberg punishment is a simple extension over the Stackelberg equilibrium solution.
In particular, it finds the point on the line segments that maximizes the leader's value subject to the follower's value being below $\theta$. For lines that fall completely below $\theta$ in terms of their follower values, we need only check the endpoints to see which is largest for the leader. For lines that span $\theta$ in terms of their follower values, we need to check the intersection point with $\theta$ as well as the endpoint that falls below $\theta$. This calculation does not increase the overall complexity over that of computing the Stackelberg equilibrium.

\section{Other Models}

The game representation in which transitions form a general graph, payoffs can occur at any node, and actions can have stochastic effects has also been called a simple stochastic game~\citep{condon92} or an alternating Markov game~\citep{littman96c}. The difference between a stochastic game~\citep{Shapley53} and an alternating stochastic game is that actions are selected non-simultaneously in an alternating stochastic game.

Since a Stackelberg equilibrium is a Stackelberg punishment with $\theta = \infty$, computing a Stackelberg punishment is at least as hard as computing a Stackelberg equilibrium. Letchford and Conitzer~\cite{letchford10} provide complexity results for a variety of Stackelberg equilibrium problems, showing that allowing stochastic transitions, simultaneous actions, or DAG-structured transition functions results in an NP-hard problem. As such, we should not expect efficient algorithms for computing Stackelberg punishment in these other game models.


\section{User Study}

\begin{figure}
    \centering
    \includegraphics[width=\columnwidth]{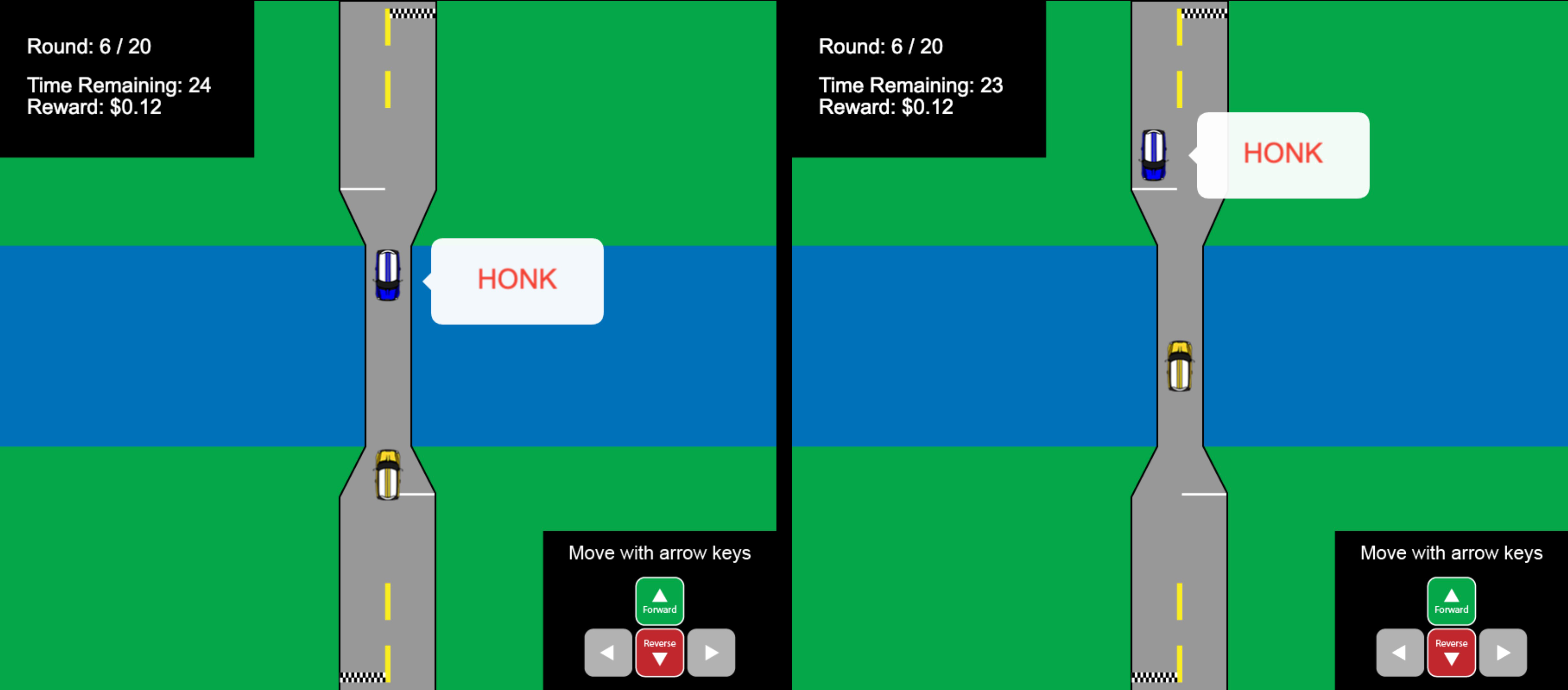}
    \caption{The SDC uses its horn to indicate its internal state. For example, in this sequence, the human driver forces the SDC (in cooperative mode) off the back of the bridge. The SDC honks to indicate that it considers itself to have been bullied and will retaliate on the next round.}
    \label{f:honk}
\end{figure}

Our motivation for studying Stackelberg punishment is as a component of an algorithm that can work productively with people. 
We conducted an experiment to assess the efficacy of this idea in a simple SDC-inspired game that requires social negotiation.

The scenario we used consists of a one-lane bridge fed from both ends by a 2-lane road (Figure~\ref{f:honk}). When two cars arrive on opposite sides of the bridge at roughly the same time, right-of-way rules dictate that the car closer to the bridge should cross, while the further car should wait. However, the further car has the opportunity to ``bully'' the closer car by crossing the bridge first, forcing the closer car to wait. In this case, a self-driving car that is hard coded to be cautious would be forced to back off the bridge, yielding to the human bully to avoid a collision, despite having the right of way.

Our experiment takes the form of an online game in which a virtual self-driving car (controlled by our algorithm) starts on one side of the bridge, displayed at the top of the screen, and a human-controlled car starts on the other side, shown at the bottom. On each turn, a player can move one position forward, stay in place, or move one position backward. The human participants control their own car's actions using the arrow keys on their keyboard.

Human participants were sourced online through Amazon Mechanical Turk and were rewarded monetarily based on how quickly they got to the other side of the bridge. The reward for each episode was \$0.13, minus \$0.01 for every two seconds before the user reached the goal. This structure was designed to encourage participants to finish quickly while still receiving a fair wage for their time (around \$15/hour). We limited our study to participants from the US to ensure that all participants had experience with similar driving laws and norms. Other demographic information was not collected.

Participants were placed into either a control group or an experimental group, and each participant completed 20 episodes of the game. At the start of each episode, one car begins noticeably closer to the bridge than the other (controlled so that each participant has an equal number of ``close'' and ``far'' starts.) We consider the closer-starting car to always have the right of way in terms of crossing the bridge first.

In response to the human participant's behavior in prior episodes, the SDC follows either a hard-coded ``cautious'' policy or a Stackelberg punishment-based policy. The policy-switching logic will be explained in Section~\ref{s:experimentalgroup}.

\subsection{Stackelberg Punishment Policy}
We computed a Stackelberg punishment policy on a simplified game with four abstract positions for each car (\emph{start}, \emph{before-bridge}, \emph{on-bridge}, \emph{finish}), resulting in a total of 16 distinct arrangements of the cars. On each decision round, one car could move forward, backward, or stay in place. We built a tree-based game over these arrangements with a maximum depth of 20 (10 decision rounds for each player). The resulting tree has 2,621,437 nodes and 1,572,862 leaves. Payoffs were computed using the same scheme as for the interactive game, \$0.13 minus \$0.01 for each step it takes to reach the finish. Each step in this abstracted game corresponds to two seconds of gameplay in the interactive game.



The result of running the algorithm on the abstracted game is shown in Figure~\ref{f:stackelberg}. It produces no more than $11$ line segments in any one node. Three behaviors for the SDC emerge, \textit{block} ($\theta < 0.10$), \textit{bully} ($0.10 \le \theta < 0.12$), and \textit{yield} ($\theta \ge 0.12$), by setting $\theta$ to different values. In the \textit{bully} case, the SDC always crosses the bridge first, regardless of starting position, as quickly as possible. This behavior is analogous to the human driver's bullying behavior. The \textit{block} strategy also takes the bridge first regardless of starting position, but drives slowly while on the bridge. Doing so decreases the reward for both players by forcing the human player to wait longer while the SDC crosses the bridge. To achieve more severe punishments, the \textit{block} strategy waits on the bridge for more time steps before proceeding to the finish line. The \textit{yield} strategy causes the SDC to let the human driver take the bridge first. For $\theta$ values other than those labeled in the figure, the SDC would behave according to a stochastic mixture of the pure strategies on either side of it.

In our experiments, we used the strategy resulting from setting $\theta=\$0.02$, which results in a Stackelberg punishment in which the SDC blocks the human driver for 9 steps (18 seconds) before proceeding. Note that the Stackelberg equilibrium strategy is for the SDC to always bully (maximizing the leader's payoff), and the minimax punishment for the game is to block the human car indefinitely (minimizing the follower's payoff). Our Stackelberg punishment strategy strikes a more humane balance between these extremes.

\subsection{Control Group}

In the control group, the SDC is controlled by a na\"{\i}ve, cautious policy: If the SDC starts farther away from the bridge than does the human driver, it will wait until the human driver passes the bridge before proceeding. If the SDC starts closer to the bridge, it will try to cross the bridge, but will back off to avoid a collision if the human driver takes the bridge. We say the human has ``bullied'' the SDC if either (1) the human forces the SDC to back off the bridge and finishes first on a round where the SDC had the right of way, or (2) if the human blocks the SDC from finishing within the round time limit (26 seconds). We hypothesized that once participants in the control group discover that they can force the SDC to yield to them, they will bully the SDC at every opportunity to maximize their monetary reward.

\subsection{Experimental Group}
\label{s:experimentalgroup}

In the experimental group, the SDC is controlled by a policy that can be in one of two driving modes, determined by a computational version of the folk theorem~\citep{littman05a,munoz08}. In \emph{cooperative} mode, the SDC is hard coded to follow right-of-way rules and avoid collisions (as in the control group). In \emph{punishing} mode, the SDC selects actions according to a computed Stackelberg punishment policy that limits the human driver's reward. Informally, the resulting policy is: go to the start of the bridge and drive forward slowly (to block the human driver) until enough time has passed that the participant's final reward cannot be above the imposed limit ($\theta=\$0.02$), then finish crossing the bridge.

We also use a horn to signal the SDC's state to the human driver. In cooperative mode, the SDC will honk while it is being bullied. In punishing mode, the SDC will honk the entire round (Figure~\ref{f:honk}). Anecdotally, we found honking to be an important signaling device. Without it, the human participants did not understand the motivation behind the SDC's reactive behavior. To our knowledge, this work is the first research that explores the use of the horn as a social signaling device for autonomous vehicles. Further experimentation is necessary to decorrelate the effects of honking from the effects of the adaptive policy, but exit survey responses (discussed in the Results section) suggest that participants' decision-making was mainly affected by the adaptive policy.

The SDC selects its mode based on the human driver's behavior in the previous round, tit-for-tat style. If the human driver obeys right-of-way rules, the SDC uses its cooperative mode in the following round. If the participant bullies the SDC, it switches to punishing mode in the following round.  This tit-for-tat strategy provides the necessary incentives to cooperate with the SDC. Other response strategies could be used, but this is left to future work.
We hypothesized that participants in the experimental group who bully the SDC at first will learn to treat the SDC fairly over the course of multiple episodes, as bullying will cause our adaptive policy to restrict the participant's subsequent reward.

To test this hypothesis, we compared occurrences of bullying between a control group of 18 participants and an experimental group of 37 participants. (We assigned fewer participants to the control group because pilot testing suggested that their behavior would have lower variability than the experimental group).

\begin{figure}
\centering
\begin{minipage}{.48\textwidth}
  \centering
    \includegraphics[width=\linewidth]{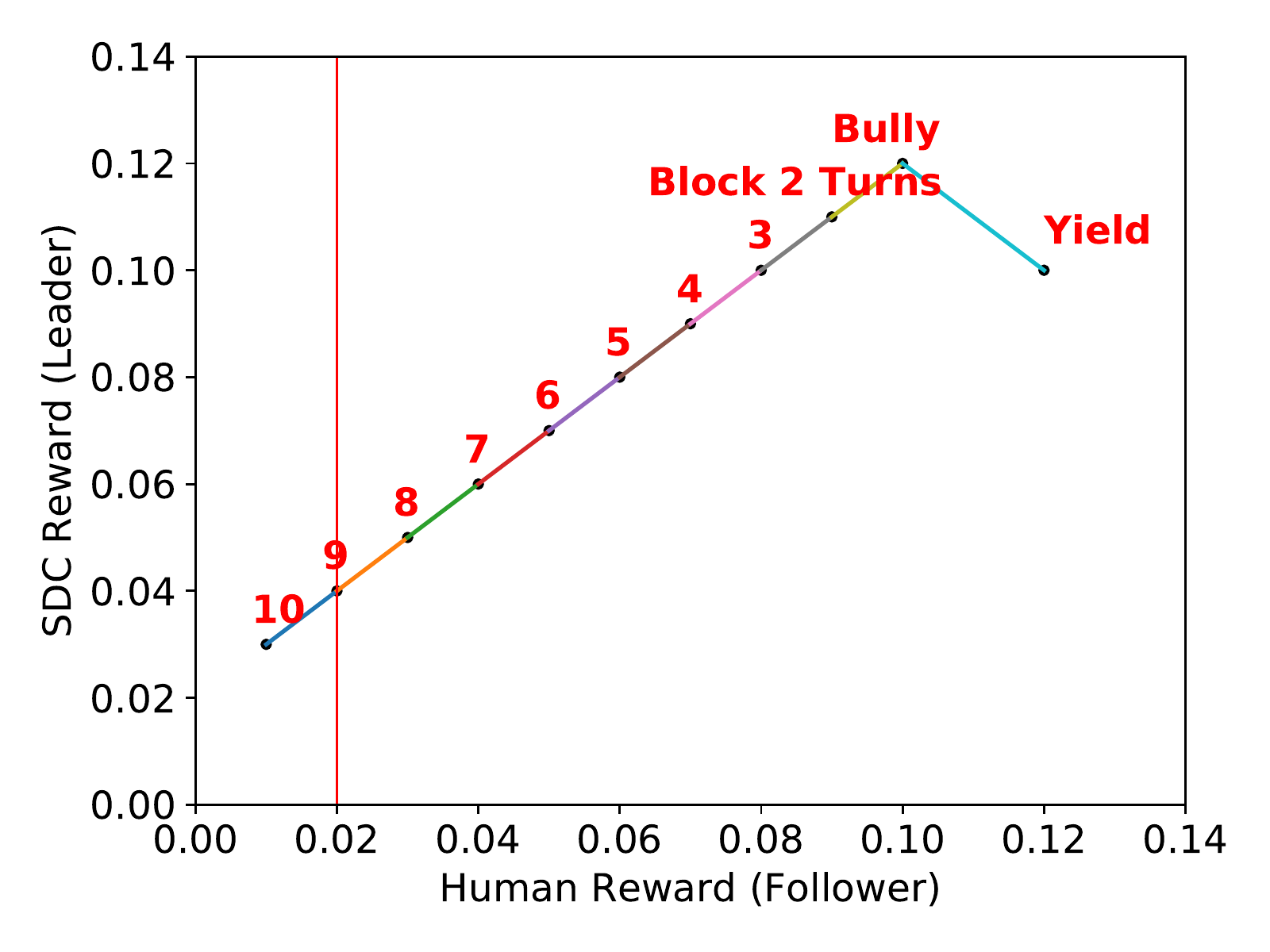}
    \caption{The Stackelberg punishment payoffs for our one-lane bridge game. Labels are a verbal description of the policy at that point. Different line colors indicate that they are separate line segments. The red vertical line indicates a $\theta$ of $\$0.02$.}
  \label{f:stackelberg}
\end{minipage}%
\hfill
\begin{minipage}{.48\textwidth}
  \centering
    \includegraphics[width=\linewidth]{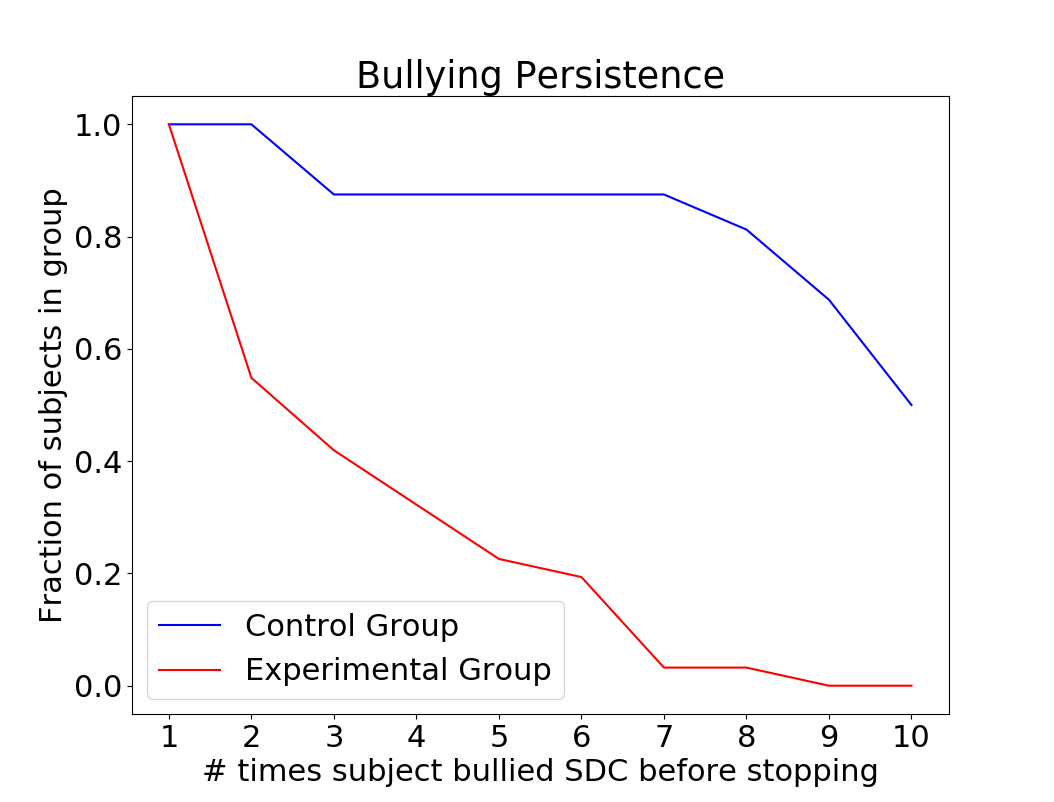}
  \caption{The relative dropoff in the number of times participants bullied the SDC, by group. Participants in the experimental group bullied far fewer times before stopping, signaling that our Stackelberg punishment policy effectively encourages drivers to behave fairly.}
    \label{f:bully-persistence}
\end{minipage}
\end{figure}

\subsection{Results}


In both the experimental and control groups, around 15\% of participants never bullied. Since the conditions look exactly the same up until the first occurrence of bullying (punishing mode is never triggered), we only consider data from participants in both groups who bullied at least once, leaving 31 and 16 participants in the experimental and control groups, respectively.

Of these participants, Figure~\ref{f:bully-persistence} shows the dropoff in the fraction who bully more than a given number of rounds. Most participants in the experimental group stop bullying after just a few initial rounds in which they experience punishment, while participants in the control group bully many more times.

The first takeaway from the control group is that human bullying of SDCs does occur. Once participants in the control group realized that the SDC would yield to them even when they did not have the right of way, they tended to take advantage of that fact at every opportunity, despite understanding it was unfair. In a post-experiment survey, control group participants commented:
\begin{quote}
\it Once I realized that the other car would reverse as soon as I crossed the line, I used it to my advantage. I would go no matter what so that I could cross the finish line faster.
\end{quote}

\begin{quote}
\it Since the other car was completely submissive, I just did whatever was in my own best interests to `win' the game.
\end{quote}

In the post-experiment survey for the experimental group, participants expressed that the adaptive policy stopped them from bullying:
\begin{quote}
\it At first it made me more aggressive, since I noticed I could easily barge my way through to get a bit of extra cash. However it only took one time for me to realize anything I gained by doing that was quickly lost in the next round as the car went agonizingly slow.
\end{quote}

When asked to rate the fairness of their driving compared to the SDC's, only 32\% of the control group described their own behavior as fair, while 91\% described the SDC's behavior as fair. In contrast, in the experimental group, 73\% of subjects described their own behavior as fair (different from the control group at $p<.005$), and 85\% described the SDC's behavior as fair (not significantly different from the control group).

It is worth noting that the reason the control group fraction in Figure~\ref{f:bully-persistence} eventually dips is because there are a limited number of rounds per subject (20 rounds), and it usually takes subjects a few rounds to ``discover'' that bullying is possible (that is, that the SDC will back off the bridge to let them pass). We expect that if the number of rounds were considerably larger, the fraction of control group participants who bully would remain high for an indefinite number of rounds, and the experimental group would drop to zero.

To quantitatively evaluate the results, we looked at how the adaptive policy influences drivers after their first exposure to the punishing mode. Participants in the experimental group face an SDC in punishing mode in the round immediately following their first occurrence of bullying, so we compared the fraction of subjects in both groups that bully only once to the fraction that bully more than once. We use a Fisher Exact Test with an alpha level of $0.05$ to determine statistical significance. Table~\ref{tab:contingency} shows the categorical data from our experiments. The result of the Fisher Exact Test gives a $p$ value of 0.0016, meaning that the adaptive Stackelberg punishment policy significantly reduced repeat bullying.

\begin{table}
\caption{The contingency table for bullying as a function of participant group.}
\label{tab:contingency}
\centering
\begin{tabular}{lrr}
 & \textbf{Control} & \textbf{Experimental} \\
\hline
\textbf{Bullied Only Once} & 0 & 14\\
\textbf{Bullied More Than Once} & 16 & 17\\
\hline
\end{tabular}
\end{table}

\section{Conclusion}

Research on self-driving cars has historically focused on the hard technical problems of perception, planning and control. Social interaction between autonomous vehicles and human drivers has been largely overlooked, but has major implications for the mainstream adoption of self-driving technology.

In this paper, we explored ``right-of-way bullying''---a social problem that could hinder the effectiveness of self-driving cars. Through an online experiment with human subjects, we showed that such bullying does occur in a simplified driving scenario. By adopting an adaptive driving policy based on a novel Stackelberg punishment formulation, we showed how to significantly decrease repeat occurrences of bullying and encourage pro-social driving behavior.

Future work should explore how
Stackelberg punishment could interact with hard-coded safety features in a production self-driving car. In addition, the solution algorithm needs to be made considerably more efficient to scale to 
more complex social behaviors with finer-grained states and actions. 

We hope that this work can be a foundation for further investigation of autonomous driving as an inherently social problem that necessitates novel technological, behavioral and sociological solutions.

\bibliographystyle{splncs04}
\bibliography{mlittman}

\end{document}